\documentclass[journal=jacsat,manuscript=article]{achemso}

\usepackage[version=3]{mhchem} % Formula subscripts using \ce{}

\author{Wenya Zhai}
\affiliation{Guangdong Provincial Key Laboratory of Magnetoelectric Physics and Devices, School of Physics, Sun Yat-sen University, Guangzhou 510275, China}
\altaffiliation{These authors contributed equally to this work.}

\author{Tingfeng Zhang}
\affiliation{Hefei National Research Center for Physical Sciences at the Microscale, CAS Key Laboratory of Strongly-Coupled Quantum Matter Physics, Department of Physics, University of Science and Technology of China, Hefei, Anhui 230026, China}
\altaffiliation{These authors contributed equally to this work.}

\author{Fengkun Chen}
\email{fkchen@dhu.edu.cn}
\affiliation{State Key Laboratory of Advanced Fiber Materials, College of Materials Science and Engineering, Donghua University, Shanghai 201620, China}
\altaffiliation{These authors contributed equally to this work.}

\author{Xiuqin Lu}
\affiliation{State Key Laboratory of Advanced Fiber Materials, College of Materials Science and Engineering, Donghua University, Shanghai 201620, China}

\author{Yunlong Xia}
\affiliation{Hefei National Research Center for Physical Sciences at the Microscale, CAS Key Laboratory of Strongly-Coupled Quantum Matter Physics, Department of Physics, University of Science and Technology of China, Hefei, Anhui 230026, China}

\author{Zengfu Ou}
\affiliation{College of Physics and Electronic Information Engineering, Guilin University of Technology, Guilin 541004, China}

\author{Ye Chen}
\affiliation{Guangdong Provincial Key Laboratory of Magnetoelectric Physics and Devices, School of Physics, Sun Yat-sen University, Guangzhou 510275, China}

\author{Donghui Guo}
\email{guodonghui@mail.sysu.edu.cn}
\affiliation{Guangdong Provincial Key Laboratory of Magnetoelectric Physics and Devices, School of Physics, Sun Yat-sen University, Guangzhou 510275, China}

\author{Meifang Zhu}
\affiliation{State Key Laboratory of Advanced Fiber Materials, College of Materials Science and Engineering, Donghua University, Shanghai 201620, China}

\author{Zhengfei Wang}
\email{zfwang15@ustc.edu.cn}
\affiliation{Hefei National Research Center for Physical Sciences at the Microscale, CAS Key Laboratory of Strongly-Coupled Quantum Matter Physics, Department of Physics, University of Science and Technology of China, Hefei, Anhui 230026, China}
\alsoaffiliation{Hefei National Laboratory, University of Science and Technology of China, Hefei, Anhui 230088, China}

\author{Jingcheng Li}
\email{lijch73@mail.sysu.edu.cn}
\affiliation{Guangdong Provincial Key Laboratory of Magnetoelectric Physics and Devices, School of Physics, Sun Yat-sen University, Guangzhou 510275, China}

\title[An \textsf{achemso} demo]
  {Tunable Flat Bands and magnetism in Triangulene-based Superatomic Graphene}

\abbreviations{IR,NMR,UV}

\begin{document}

%%%%%%%%%%%%%%%%%%%%%%%%%%%%%%%%%%%%%%%%%%%%%%%%%%%%%%%%%%%%%%%%%%%%%
%% The "tocentry" environment can be used to create an entry for the
%% graphical table of contents. It is given here as some journals
%% require that it is printed as part of the abstract page. It will
%% be automatically moved as appropriate.
%%%%%%%%%%%%%%%%%%%%%%%%%%%%%%%%%%%%%%%%%%%%%%%%%%%%%%%%%%%%%%%%%%%%%

%%%%%%%%%%%%%%%%%%%%%%%%%%%%%%%%%%%%%%%%%%%%%%%%%%%%%%%%%%%%%%%%%%%%%
%% The abstract environment will automatically gobble the contents
%% if an abstract is not used by the target journal.
%%%%%%%%%%%%%%%%%%%%%%%%%%%%%%%%%%%%%%%%%%%%%%%%%%%%%%%%%%%%%%%%%%%%%
\begin{abstract}
Superatomic graphene platforms host a rich portfolio of flat-band-driven exotic quantum properties, yet their experimental realization remains challenging. Here, we report the bottom-up on-surface synthesis of superatomic graphene using phosphorus-doped triangulene as building blocks. Scanning tunneling microscopy and spectroscopy measurements resolve the well-defined honeycomb lattice of as-fabricated superatomic graphene and demonstrate the characteristic Dirac band and flat band electronic structures. Density functional theory calculations reveal that the flat bands originate from the in-plane p$_x,_y$-like frontier orbitals of the phosphorus-doped triangulene units, leading to intrinsic half-metallic behavior. Furthermore, oxygen functionalization of the molecular precursor enables deterministic modulation of the electronic structure and magnetic ordering. This work establishes a general platform for designing correlated quantum materials with tunable flat band properties.
\end{abstract}

%%%%%%%%%%%%%%%%%%%%%%%%%%%%%%%%%%%%%%%%%%%%%%%%%%%%%%%%%%%%%%%%%%%%%
%% Start the main part of the manuscript here.
%%%%%%%%%%%%%%%%%%%%%%%%%%%%%%%%%%%%%%%%%%%%%%%%%%%%%%%%%%%%%%%%%%%%%
\section{Introduction}
In flat-band materials, the strong electron correlation effects induced by the narrow bandwidth, in conjunction with the highly accumulated density of states, can lead to a variety of exotic quantum properties, including ferromagnetism\cite{Tasaki1998}, high-temperature superconductivity\cite{Miyahara,Peotta2015}, and fractional quantum Hall effects\cite{Sheng2011,PBernevig2011,Neupert2011}. Theoretical studies on the intrinsic connection between flat bands and itinerant ferromagnetism can be traced back to pioneering studies in the early 1990s\cite{AMielke_1991,AMielke_1991-2,AMielke_1992}, which demonstrated that Hubbard models defined on a broad class of line graphs can exhibit ferromagnetic ground states\cite{Tasaki1998,Liu2014}. These flat-band-induced magnetic behaviors have been directly validated by recent experimental studies in kagome lattices\cite{Yin2022}, and further extended to other systems such as zigzag graphene nanoribbons\cite{Son2006,Ruffieux2016,Blackwell2021,Li2021} and electrically gated twisted bilayer graphene\cite{Sharpe2019}.

Beyond lattice geometry, flat bands (FBs) can also be realized by harnessing orbital degrees of freedom. For instance, in superatomic graphene, replacing the out-of-plane \emph{p}$_z$ orbital at each honeycomb lattice site with the in-plane \emph{p}$_x,_y$-orbital of the superatom (molecule),  enables the creation of flat bands spanning the entire Brillouin zone \cite{Wu2008}. Such flat bands host a rich variety of correlated phenomena, which can ultimately give rise to ferromagnets\cite{Anindya2022,Yu2026}, half-metals\cite{Kan2012,Anindya2022}, Mott insulators\cite{Zhou2021,Anindya2022}, excitonic insulators\cite{Sethi2021} and so on\cite{Jing2019}. Moreover, chemical modification or heteroatom doping of the superatomic constituents enables precise control over the flat-band filling and exchange interactions, thereby providing a feasible route toward deterministic manipulation of magnetic ordering and correlated electronic behavior\cite{Yu2024}. Despite these exciting predictions, the experimental realization of superatomic graphene with tunable flat bands and controllable magnetic properties remains highly challenging.

Here, we report the bottom-up fabrication of triangulene-based superatomic graphene with tunable flat-band electronic structures and emergent magnetic properties through a combined scanning tunneling microscopy (STM) measurements and density functional theory (DFT) simulations. The as-synthesized superatomic graphene is fabricated via an on-surface synthetic strategy\cite{Talirz2016,Galeotti2020,Delgado2023} using phosphorus-doped triangulene and its oxygen-functionalized derivative as monomer building blocks. STM measurements resolve the precise lattice structure of the as-fabricated superatomic graphene, identifying their flat band electronic structures.  Complementary DFT simulations quantitatively reproduce the experimentally observed band characteristics, uncovering the origin of spin-polarized bands from the \emph{p}$_x,_y$-like orbital of superatomic graphene. 

\section{Results and discussion}

\begin{figure}[!htb]
\centering
\includegraphics[width=0.9\textwidth]{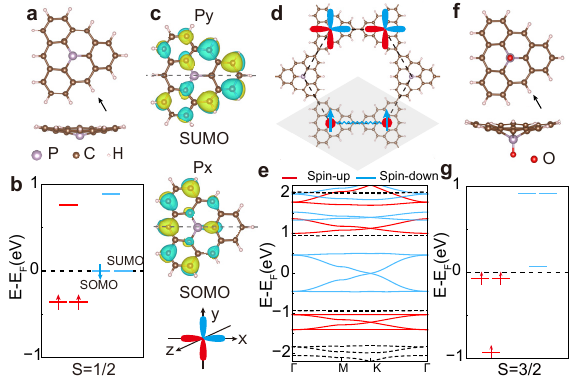}
\caption{\textbf{Electronic and magnetic properties of triangulene based superatomic graphene}. 
\textbf{a}, Model structure of PT molecule from top and side views (arrow indicates the side view direction). 
\textbf{b}, Energy levels of PT, showing doublet ground states (S=1/2). 
\textbf{c}, DFT simulated FMOs of PT, which can be considered as \emph{p}$_x$, \emph{p}$_y$-orbitals.  
\textbf{d}, Model structure of PTSG. The dashed hexagon highlight the formed honeycomb lattice. Each sites of the lattice are occupied by PT, with its \emph{p}$_x,_y$ orbital superimposed on the image. The gray rhombus displays the primary unit cell with two interacting spins from each PT molecule.
\textbf{e}, Simulated band structures of PTSG (solid lines) and OPTSG (dashed lines). The red (light blue) curves indicating the spin-up (spin down) band structures, while the dashed lines indicate the anti-ferromagnetic ground states of OPTSG.
\textbf{f}, Model structure of OPTSG from top and side views (arrow indicates the side view direction). 
\textbf{g}, Energy levels of PT molecule oxide, showing Quartet ground states (S=3/2).}
\end{figure}

Fig. 1a depicts the model structure of phosphorus-doped triangulene(PT). DFT simulations reveal that freestanding PT adopts a nonplanar configuration, which gives rise to a doublet ground state (Fig. 1b). Its frontier molecular orbitals (FMOs) are predominantly localized at the edge carbon atoms of the triangulene backbone, which can be considered as the \emph{p}$_x$, \emph{p}$_y$-orbitals (Fig. 1c). Replacing each lattice site of a graphene honeycomb lattice with a PT molecule yields a PT-based superatomic graphene (PTSG) framework, as shown in Fig. 1d. In this periodic architecture, the FMOs of individual PT units act analogously to the in-plane \emph{p}$_x,_y$ orbital of the honeycomb lattice (Fig. 1c, d), giving rise to the generation of flat bands. The calculated spin-polarized band structure of PTSG (Fig. 1e) exhibits multiple well-defined flat bands near the Fermi level. Notably, the PTSG framework exhibits intrinsic half-metallic behavior. The spin-down electron band occupies the Fermi level, featuring two flat bands bridged by two Dirac bands, while there is a band gap of around 2 eV for the spin-up electron band. 

Precise chemical modification of the triangulene building blocks enables fine-tailoring of the flat band characteristics and associated magnetic properties of the PTSG lattice. For instance, covalent attachment of an oxygen atom to the phosphorus dopant of the PT unit switches the ground state of the triangulene moiety from a doublet (S=1/2) to a quartet (S=3/2) (Figs. 1f, g). In this spin configuration, two spin-up electrons occupy the states close to Fermi level, thereby changing the interunit magnetic coupling between neighboring triangulene moieties from ferromagnetic to antiferromagnetic. Consequently, the oxygen-functionalized PTSG (OPTSG) framework undergoes substantial electronic reconstruction. The flat bands shift away from the Fermi level (dashed black curves in Fig. 1e), accompanied by the opening of a band gap of approximately 1.86 eV. These electronic changes are accompanied by a transition in the  magnetic order of the framework, from half-metallic ferromagnetic to semiconducting antiferromagnetic.

\begin{figure}[!b]
\centering
\includegraphics[width=\textwidth]{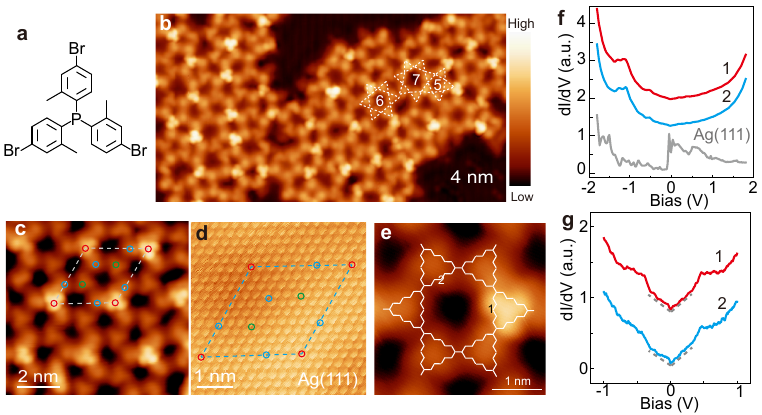}
\caption{\textbf{Electronic and magnetic properties of PTSG}. 
\textbf{a}, Chemical structure of molecular precursor tris(4-bromo-2-methylphenyl) phosphine. 
\textbf{b}, STM overview image of the formed PTSG on Ag(111) surface (V = 1 V). The formed 5-, 6-, and 7-membered rings are indicated in the figure, with each dashed triangular representing a PT molecule. 
\textbf{c}, STM image of an area with unitary 6-membered rings, displaying the periodicity of the bright PT molecule (dashed rhombus). Colored circles highlight the phosphorus atoms of PT in the unit cell.
\textbf{d}, STM image of Ag(111) surface with atomic resolution, showing the proposed adsorption sites of PTSG. The dashed rhombus share the same size as the rhombus in g. Red circles denote phosphorus atoms above Ag atoms, while light blue circles indicate placement of phosphorus atoms on bridge or hollow sites.
\textbf{e}, Zoom-in STM image of PTSG with its contour superimposed. 
\textbf{f, g}, Wide and narrow range d\textit{I}/d\textit{V} spectra taken on PTSG. (spectra condition: V = 1.8 V, I = 300 pA for spectra in b, and V = 1 V, I = 300 pA  for spectra in c). The locations of spectra are noted in \textbf{e} with numbers. The gray dashed lines highlight the V-shaped feature.}
\end{figure}

The molecular precursor tris(4-bromo-2-methylphenyl) phosphine (Fig. 2a) was designed for the on-surface synthesis of PTSG through Ullmann coupling and cyclodehydrogenation. Similar reaction path ways have previously been employed for the synthesis of triangulene-based graphene structures\cite{Mishra2021c,Delgado2023,Yan2026}. The molecular precursors were sublimed onto a clean Ag(111) substrate at room temperature, followed by stepwise annealing at 140 $^{\circ}$C and 200 $^{\circ}$C to activate the on-surface reactions (Supplementary Note. 2). Fig. 2b shows a scanning tunneling microscopy (STM) image of the formed structure with honeycomb lattice characteristics, together with a certain fraction of non-6-membered rings. Statistics from a half-coverage sample show an average domain size of about 35 by 35 nm$^2$, with 5-, 6-, and 7-membered rings accounting for
27.0\%, 53.3\% and 15.2\%, respectively (Supplementary Fig. 9). The ordered structure consisting of unitary 6-membered rings can reach sizes of about 8 by 8 nm$^2$ (Fig. 2c). Within these ordered regions, one-sixth of the PT molecules appear brighter than others, giving rise to an additional superstructure periodicity. The distance between two adjacent bright PT molecules is 2.88 nm, a value exactly 10 times the lattice constant of Ag(111). We speculate that bright PT molecules absorb with P atoms at the top sites, while others absorb with P atoms at the hollow or bridge sites (Fig. 2d). The ratio between the top sites and the hollow or bridge locations is 1 to 5 in each unit cell (dashed rhombus in Fig. 2d), indicating the preferred adsorption configuration. 

To study the electronic properties of PTSG, differential conductance spectra (d\textit{I}/d\textit{V}) were recorded on both bright and normal molecules (Fig. 2f,g). The spectra exhibit identical double peaks near -1.2 V, along with sharp increases at $\pm$ 1.7 V. In addition, less distinct features close to the Fermi level appear in the spectra. These features are better resolved in the narrow range spectra, displaying a V-shaped  line profile centered at the Fermi level (Fig. 2g). Since the d\textit{I}/d\textit{V} spectrum is a measure of the local density of states (LDOS), the absence of a band gap around the Fermi level indicates the metallic behavior of PTSG. Notably, the observed V-shaped line profile is highly consistent with the simulated Dirac band structure near the Fermi level. It is worth noting that defects (non 6-membered rings) have no apparent effect on the electronic  properties of PTSG (Supplementary Note. 6). As shown in the Supplementary Fig. 15, spectrum taken on PT molecules in these rings displays similar V-shaped feature as the spectra from PT molecules in ordered structures.

\begin{figure} [!t]
\centering
\includegraphics[width=0.9\textwidth]{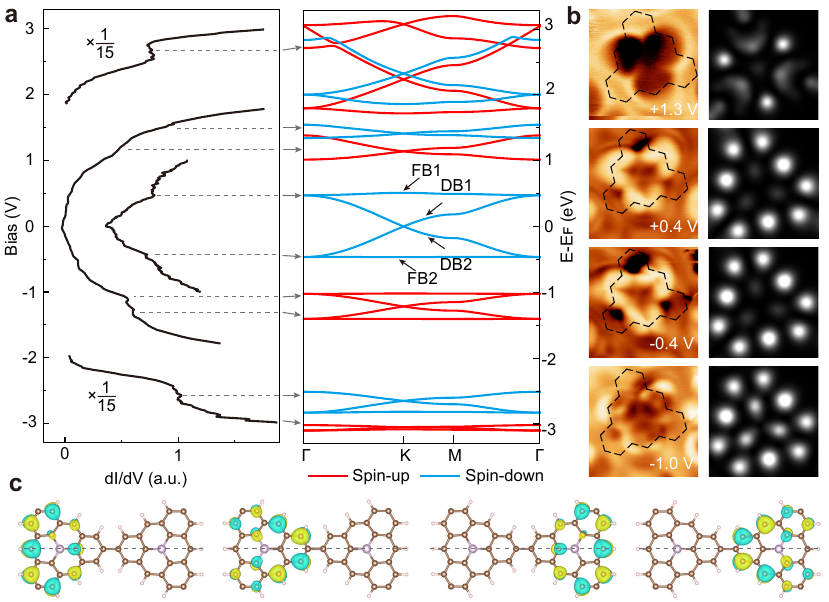}
\caption{\textbf{Theoretical electronic structures of PTSG}. 
\textbf{a}, Comparison between measured d\textit{I}/d\textit{V} spectra and calculated band structures(set point for spectra in range ($\pm$2 V, $\pm$3 V): V = 3 V, I = 100 pA, the other two spectra are the same spectra 2 in Fig. 2b, c). The flat bands and Dirac bands around Fermi level are labeled as FB1, FB2, DB1, DB2 respectively. 
\textbf{b}, Comparison between measured STM d\textit{I}/d\textit{V} maps and simulated maps at bias values as noted in the figures. The STM maps are obtained with a CO-terminated tip (image size: 1 x 1 nm$^2$, map condition: I = 50 pA).   
\textbf{c}, Wannier orbitals as the basis functions for the four-band structure around the Fermi level.  
 }
\end{figure}

Fig. 3a correlates the experimentally measured d\textit{I}/d\textit{V} spectra with the calculated spin-polarized band structures of PTSG. The calculations reveal that the spin-down electron band occupies the Fermi level, featuring two flat bands bridged by two Dirac bands (named FB1, FB2, DB1, DB2). The energetic positions of these four bands align well with the V-shaped feature observed in the experimental d\textit{I}/d\textit{V} spectra (indicated by dashed lines in Fig. 3a). In constrast, there is a band gap exceeding 2 eV for spin-up electrons, with the conduction band minimum near 1 eV and upper valence bands (four band characteristics) ranging from -1 eV to -1.4 eV. Close to these energy locations, a step at 1.2 V and double peaks at -1.1 V, -1.2 V appear on the spectra, indicating their origin from the spin-up electronic bands. d\textit{I}/d\textit{V} maps at close energies (1.3 V, $\pm$ 0.4 V, -1 V) show three-fold symmetries with enhancement at the edge of PT molecules, which are reproduced by the simulated maps at the same energies (Fig. 3b). Additionally, the sharp conductance onsets observed at higher bias energies (around $\pm 2.5 V$ and $\pm 2.9 V$) in experimental d\textit{I}/d\textit{V} spectra also correspond to flat band features in the simulated band structures (indicated by dashed lines in Fig. 3a).The excellent agreement between experimental observations and DFT simulations rationalizes the observed electronic properties of PTSG, further confirms its half-metallic character.

To further uncover the origin of the four bands (FB1, FB2, DB1, and DB2) occupying the Fermi level, the Wannier functions for the four bands are derived (Fig. 3c). The resulting Wannier functions exhibit characteristic features as singly occupied molecular orbital (SOMO) and singly unoccupied molecular orbital (SUMO) of PT molecules(Fig.1c), indicating that the four bands originate from the hybridization of the frontier orbitals of neighboring triangulene superactoms. This finding provides direct experimental support for flat band engineering through the theoretic model of  \emph{p}$_x,_y$-orbital honeycomb lattice\cite{Wu2008}.  

\begin{figure}[!htb]
\centering
\includegraphics[width=0.7\textwidth]{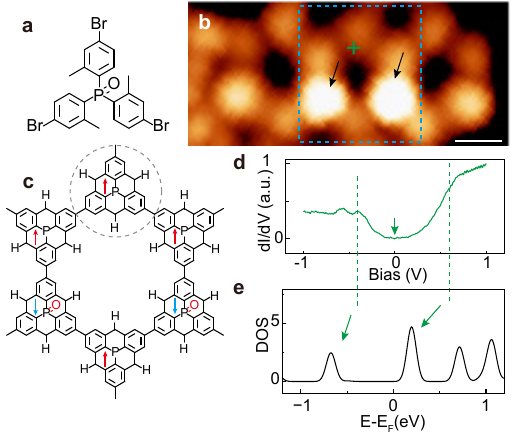}
\caption{\textbf{Synthesis and characterzation of O-PTSG}. 
\textbf{a}, chemical structure of molecular precursor tris(4-bromo-2-methylphenyl) phosphine oxide.
\textbf{b},STM image of the formed PTSG oxide on Ag(111) surface (V = 1 V, I = 50 pA). 
\textbf{c}, Chemical structures of the dashed area in b. The arrows in red and light blue indicate the spin configuration of this structures from DFT simulations. 
\textbf{d}, d\textit{I}/d\textit{V} spectra taken on PT molecules with location indicated in b (spectra condition: V = 1 V, I = 1 nA).
\textbf{e}, Simulated Density of states (DOS) at the same  location of spectra in d.
}
\end{figure}

To further tune the flat band structures of PTSG, the molecular precursor used to synthesize PTSG is modified. An oxygen atom is covalently attached to the phosphorus center, yielding tris (4-bromo-2-methylphenyl) phosphine oxide as the functionalized building block. Oxygen functionalization switches the spin ground state of the PT from S = 1/2 to S = 3/2 (Fig. 1g), resulting in a fundamentally altered band structure for the O-PTSG framework(Fig. 1e). O-PTSG was fabricated via an on-surface reaction protocol analogous to that employed for pristine PTSG. A representative STM image resolves the formed O-PTSG structures on Ag (111) substrate (Fig. 4b), displaying a well-defined honeycomb lattice with bright protrusions at randomly distributed lattice sites(indicated by arrows in Fig. 4b). These protrusions are assigned to covalently bound oxygen atoms, as shown in the structural model in Fig. 4c. Quantitative analysis from multiple STM images indicates that only ~8\% of the oxygen moieties remain after the high-temperature annealing step required for on-surface reactions. The d\textit{I}/d\textit{V} spectrum taken at the edges of PT molecules adjacent to oxygen-modified sites exhibits a prominent, flat zero-conductance region centered at the Fermi level (indicated by arrow in Fig. 4d).  Two well-resolved conductance step features appear at -0.41 V and 0.60 V (marked by dashed lines), indicating an experimental band gap of 1.01 eV. These results demonstrate that even a low concentration of oxygen functionalization is sufficient to induce substantial reconstruction of the flat-band electronic structure of PTSG. 

 DFT simulations were performed to elucidate the effect of oxygen functionalization on spin ordering and electronic structure. The calculation resutls show that spins in unmodified PT units remain ferromagnetically aligned (marked by red arrows in Fig. 4c), while spins in oxygen-bonded PT units adopt antiferromagnetic coupling relative to their unmodified neighbors. The DFT-simulated LDOS for the PT unit highlighted by the gray dashed circle in Fig. 4c (the same site where the experimental STS spectrum was recorded) exhibits two distinct peaks near the Fermi level, located at -0.68 eV and 0.20 eV respectively. The 0.88 eV energy separation between these peaks is in good agreement with the experimentally measured band gap of 1.01 eV, corroborating the semiconducting behavior of OPTSG. The simulations are consistent with the experimentally observed transition from half-metallic to semiconducting behavior, demonstrating the role of oxygen functionalization on the band structures and spin orders .

In summary, we have successfully fabricated two types of triangulene-based superatomic graphene via a bottom-up on-surface synthetic strategy. Their tunable flat bands are identified through a combination of STM measurements and DFT simulations. This work provides experimental insights into the \emph{p}$_x,_y$-orbital model of the honeycomb lattice and establishes a versatile platform for the development of 2D carbon magnets with tailored magnetic properties. Based on theoretical predictions\cite{Jing2019,Zhou2021,Sethi2021,Anindya2022,Yu2024,Yu2026} and our experimental results, we foresee that 2D carbon magnets with desired electronic and magnetic properties could be synthesized, which may find applications in reducing the size of spintronic devices with improved performance.  

\section*{Methods}

\subsection{Synthesis of molecular precursors}	
Tris(4-bromo-2-methylphenyl) phosphane was synthesized by following the reported procedure\cite{Trunk2017}.

\subsection{Sample preparation and experimental details.}
The experiments were performed on a commercial low-temperature ultra-high-vacuum (UHV) STM (Unisoku USM 1300). The Ag(111) single crystal (MaTecK GmbH, 99.999$\%$) was cleaned via repeated cycles of Ar$^+$ sputtering and subsequent annealing at 550 \textcelsius. The molecular precursors were sublimed from Knudsen cells onto a clean Ag(111) substrate kept at room temperature.
The sublimation temperature of the molecular precursors was {85} $^{\circ}$C. The sample was then step-wisely annealed at 140 $^{\circ}$C and 200 $^{\circ}$C (10 minutes for each step). The STM/STS measurements were performed at 5 K and 0.4 K. A tungsten tip was used for images and spectroscopy, with the use of a CO-terminated tip noted in the paper. The d\emph{I}/d\emph{V} signals were recorded using a lock-in amplifier with a bias modulation of \emph{V}$\rm_{rms}$ = 10 mV (Fig. {2b, 2c, 3a, 3b}), and \emph{V}$\rm_{rms}$ = 0.5 mV (Fig. {2d, 2e, 2g, 4c}) at a modulation frequency of 479 Hz. All the STM images were processed with WSxM\cite{Horcas2007}.

\subsection{DFT simulations.}	

The first-principles calculations were performed to determine the electronic properties using the Vienna Ab initio Simulation Package (VASP)\cite{Kresse1996} with the Perdew–Burke–Ernzerhof functional\cite{Perdew1996PRL}. The plane-wave cutoff of 520 eV. During structural relaxation, all atoms were fully optimized with a force convergence criterion of 0.001 eV/Å and a total energy tolerance of 10$^{-6}$ eV. For the PT molecule and its oxide, a vacuum layer of 15 Å was introduced along the x, y, and z directions to eliminate interactions between periodic images. For two-dimensional crystals, a 15 Å vacuum layer was applied along the out-of-plane direction to ensure slab decoupling. The k-point mesh was set to 5×5×1 for PTSG and OPTSG, and to 1×1×1 for oxygen-doped PTSG (whose unit cell is $\sqrt3$×$\sqrt3$ times larger than that of PTSG), with all calculations performed using the PBE0 functional\cite{Perdew1996,Ernzerhof1999,Adamo1999}. Based on the Tersoff–Hamann approximation\cite{Tersoff1985}, scanning tunneling spectroscopy images were simulated from the calculated local density of states (LDOS), with an increased k-point sampling density of 13×13×1 to improve accuracy. Wannier bands and orbitals were constructed using the Wannier90 package\cite{Arash2014} via maximally localized Wannier functions (MLWFs). 
%%%%%%%%%%%%%%%%%%%%%%%%%%%%%%%%%%%%%%%%%%%%%%%%%%%%%%%%%%%%%%%%%%%%%
%% The "Acknowledgement" section can be given in all manuscript
%% classes.  This should be given within the "acknowledgement"
%% environment, which will make the correct section or running title.
%%%%%%%%%%%%%%%%%%%%%%%%%%%%%%%%%%%%%%%%%%%%%%%%%%%%%%%%%%%%%%%%%%%%%
\begin{acknowledgement}

We acknowledge financial support from National Natural Science Foundation of China 
(Nos. 12474181, 12574526, 22171044, 12174369, 22171044, 12494591), Guangdong Major Project of Basic and Applied Basic Research (2021B0301030002), Guangdong Basic and Applied Basic Research (2024A1515010656), Quantum Science and Technology-National Science and Technology Major Project (No. 2021ZD0302800). The experiments reported were conducted at the Guangdong Provincial Key Laboratory of Magnetoelectric Physics and Devices, No. 2022B1212010008. We thank Supercomputing Center at USTC and Hefei Advanced Computing Center for providing computing resources.

\end{acknowledgement}

%%%%%%%%%%%%%%%%%%%%%%%%%%%%%%%%%%%%%%%%%%%%%%%%%%%%%%%%%%%%%%%%%%%%%
%% The same is true for Supporting Information, which should use the
%% suppinfo environment.
%%%%%%%%%%%%%%%%%%%%%%%%%%%%%%%%%%%%%%%%%%%%%%%%%%%%%%%%%%%%%%%%%%%%%
\begin{suppinfo}

Detailed Analysis and additional STM images are available on-line in supplementary Information.

\end{suppinfo}

%%%%%%%%%%%%%%%%%%%%%%%%%%%%%%%%%%%%%%%%%%%%%%%%%%%%%%%%%%%%%%%%%%%%%
%% The appropriate \bibliography command should be placed here.
%% Notice that the class file automatically sets \bibliographystyle
%% and also names the section correctly.
%%%%%%%%%%%%%%%%%%%%%%%%%%%%%%%%%%%%%%%%%%%%%%%%%%%%%%%%%%%%%%%%%%%%%
\bibliography{ms}

@article{Yu2026,
author = {Yu, Hongde and Heine, Thomas},
title = {Metal-Free Ferromagnetism in Triangulene Two-Dimensional Frameworks},
journal = {Journal of the American Chemical Society},
volume = {148},
number = {13},
pages = {13822-13833},
year = {2026},
}

@article{Perdew1996PRL,
  title = {Generalized Gradient Approximation Made Simple},
  author = {Perdew, John P. and Burke, Kieron and Ernzerhof, Matthias},
  journal = {Phys. Rev. Lett.},
  volume = {77},
  issue = {18},
  pages = {3865--3868},
  numpages = {0},
  year = {1996},
  month = {Oct},
  publisher = {American Physical Society},
  doi = {10.1103/PhysRevLett.77.3865},

}

@article{Arash2014,
title = {An updated version of wannier90: A tool for obtaining maximally-localised Wannier functions},
journal = {Computer Physics Communications},
volume = {185},
number = {8},
pages = {2309-2310},
year = {2014},
issn = {0010-4655},
doi = {https://doi.org/10.1016/j.cpc.2014.05.003},

author = {Arash A. Mostofi and Jonathan R. Yates and Giovanni Pizzi and Young-Su Lee and Ivo Souza and David Vanderbilt and Nicola Marzari},


}

@article{Tersoff1985,
  title = {Theory of the scanning tunneling microscope},
  author = {Tersoff, J. and Hamann, D. R.},
  journal = {Phys. Rev. B},
  volume = {31},
  issue = {2},
  pages = {805--813},
  numpages = {0},
  year = {1985},
  month = {Jan},
  publisher = {American Physical Society},
  doi = {10.1103/PhysRevB.31.805},
  url = {https://link.aps.org/doi/10.1103/PhysRevB.31.805}
}

@article{AMielke_1991,
doi = {10.1088/0305-4470/24/2/005},
year = {1991},
month = {jan},
publisher = {},
volume = {24},
number = {2},
pages = {L73},
author = {A Mielke},
title = {Ferromagnetic ground states for the Hubbard model on line graphs},
journal = {Journal of Physics A: Mathematical and General}
}

@article{AMielke_1991-2,
doi = {10.1088/0305-4470/24/14/018},
year = {1991},
month = {jul},
publisher = {},
volume = {24},
number = {14},
pages = {3311},
author = {A Mielke},
title = {Ferromagnetism in the Hubbard model on line graphs and further considerations},
journal = {Journal of Physics A: Mathematical and General}
}

@article{Miyahara,
title = {BCS theory on a flat band lattice},
journal = {Physica C: Superconductivity},
volume = {460-462},
pages = {1145-1146},
year = {2007},
note = {Proceedings of the 8th International Conference on Materials and Mechanisms of Superconductivity and High Temperature Superconductors},
issn = {0921-4534},
doi = {https://doi.org/10.1016/j.physc.2007.03.393},
author = {S. Miyahara and S. Kusuta and N. Furukawa}
}

@article{AMielke_1992,
doi = {10.1088/0305-4470/25/16/011},
year = {1992},
month = {aug},
publisher = {},
volume = {25},
number = {16},
pages = {4335},
author = {A Mielke},
title = {Exact ground states for the Hubbard model on the Kagome lattice},
journal = {Journal of Physics A: Mathematical and General}
}

@article{Tasaki1998,
    author = {Tasaki, Hal},
    title = {From Nagaoka's Ferromagnetism to Flat-Band Ferromagnetism and Beyond: An Introduction to Ferromagnetism in the Hubbard Model},
    journal = {Progress of Theoretical Physics},
    volume = {99},
    number = {4},
    pages = {489-548},
    year = {1998},
    month = {04},
    issn = {0033-068X},
    doi = {10.1143/PTP.99.489},

}

@article{Peotta2015,
author = {Peotta, Sebastiano},
doi = {10.1038/ncomms9944},
pages = {1--9},
title = {{Superfluidity in topologically nontrivial flat bands}},
year = {2015}
}

@article{Sheng2011,
author = {Sheng, D N and Gu, Zheng-cheng and Sun, Kai and Sheng, L},
doi = {10.1038/ncomms1380},
journal = {Nature Communications},
publisher = {Nature Publishing Group},
title = {{Fractional quantum Hall effect in the absence of Landau levels}},
year = {2011}
}

@article{PBernevig2011,
  title = {Fractional Chern Insulator},
  author = {Regnault, N. and Bernevig, B. Andrei},
  journal = {Phys. Rev. X},
  volume = {1},
  issue = {2},
  pages = {021014},
  numpages = {14},
  year = {2011},
  month = {Dec},
  publisher = {American Physical Society},
  doi = {10.1103/PhysRevX.1.021014},
}

@article{Neupert2011,
  title = {Fractional Quantum Hall States at Zero Magnetic Field},
  author = {Neupert, Titus and Santos, Luiz and Chamon, Claudio and Mudry, Christopher},
  journal = {Phys. Rev. Lett.},
  volume = {106},
  issue = {23},
  pages = {236804},
  numpages = {4},
  year = {2011},
  month = {Jun},
  publisher = {American Physical Society},
  doi = {10.1103/PhysRevLett.106.236804},
}

@article{Yan2026,
author = {Yan, Yuyi and Liu, Fujia and Tang, Weichen and Wong, Han Xuan and Qie, Boyu and Louie, Steven G. and Fischer, Felix R.},
doi = {10.1038/s41563-026-02528-3},
issn = {14764660},
journal = {Nature Materials},
publisher = {Springer US},
title = {{Engineering phase-frustration-induced flat bands in an aza-triangulene covalent kagome lattice}},
year = {2026}
}

@article{Yin2022,
archivePrefix = {arXiv},
arxivId = {2212.11628},
author = {Yin, Jia Xin and Lian, Biao and Hasan, M. Zahid},
doi = {10.1038/s41586-022-05516-0},
eprint = {2212.11628},
issn = {14764687},
journal = {Nature},
number = {7941},
pages = {647--657},
pmid = {36543954},
publisher = {Springer US},
title = {{Topological kagome magnets and superconductors}},
volume = {612},
year = {2022}
}

@article{Son2006,
author = {Son, Young-Woo and Cohen, Marvin L and Louie, Steven G},
doi = {10.1038/nature05180},
issn = {1476-4687},
journal = {Nature},
volume = {444},
number = {7117},
pages = {347--349},
title = {{Half-metallic graphene nanoribbons}},
year = {2006}
}

@article{Kan2012,
author = {Kan, Erjun and Hu, Wei and Xiao, Chuanyun and Lu, Ruifeng and Deng, Kaiming and Yang, Jinlong and Su, Haibin},
doi = {10.1021/ja210822c},
issn = {0002-7863},
journal = {J. Am. Chem. Soc.},
month = {apr},
number = {13},
pages = {5718--5721},
publisher = {American Chemical Society},
title = {{Half-Metallicity in Organic Single Porous Sheets}},
volume = {134},
year = {2012}
}

@article{Jing2019,
author = {Jing, Yu and Heine, Thomas},
doi = {10.1021/jacs.8b09900},
journal = {J. Am. Chem. Soc.},
number = {2},
pages = {743--747},
title = {{Two-Dimensional Kagome Lattices Made of Hetero Triangulenes Are Dirac Semimetals or Single-Band Semiconductors}},
volume = {141},
year = {2019}
}

@article{Ruffieux2016,
author = {Ruffieux, Pascal and Wang, Shiyong and Yang, Bo and S{\'{a}}nchez-S{\'{a}}nchez, Carlos and Liu, Jia and Dienel, Thomas and Talirz, Leopold and Shinde, Prashant and Pignedoli, Carlo A and Passerone, Daniele and Dumslaff, Tim and Feng, Xinliang and M{\"{u}}llen, Klaus and Fasel, Roman},
doi = {10.1038/nature17151},
journal = {Nature},
number = {7595},
pages = {489--492},
publisher = {Nature Publishing Group, a division of Macmillan Publishers Limited. All Rights Reserved.},
title = {{On-surface synthesis of graphene nanoribbons with zigzag edge topology}},
volume = {531},
year = {2016}
}

@article{Blackwell2021,
author = {Blackwell, Raymond E. and Zhao, Fangzhou and Brooks, Erin and Zhu, Junmian and Piskun, Ilya and Wang, Shenkai and Delgado, Aidan and Lee, Yea Lee and Louie, Steven G. and Fischer, Felix R.},
doi = {10.1038/s41586-021-04201-y},
issn = {14764687},
journal = {Nature},
month = {dec},
number = {7890},
pages = {647--652},
pmid = {34937899},
publisher = {Nature Research},
title = {{Spin splitting of dopant edge state in magnetic zigzag graphene nanoribbons}},
volume = {600},
year = {2021}
}

@article{Sharpe2019,
author = {Sharpe, Aaron L. and Fox, Eli J. and Barnard, Arthur W. and Finney, Joe and Watanabe, Kenji and Taniguchi, Takashi and Kastner, M. A. and Goldhaber-Gordon, David},
doi = {10.1126/science.aaw3780},
issn = {10959203},
journal = {Science},
number = {6453},
pages = {605--608},
pmid = {31346139},
title = {{Emergent ferromagnetism near three-quarters filling in twisted bilayer graphene}},
volume = {365},
year = {2019}
}

@article{Wu2008,
author = {Wu, Congjun and {Das Sarma}, S.},
doi = {10.1103/PhysRevB.77.235107},
issn = {10980121},
journal = {Phys. Rev. B},
number = {23},
pages = {23},
title = {{px,y -orbital counterpart of graphene: Cold atoms in the honeycomb optical lattice}},
volume = {77},
year = {2008}
}

@article{Zhou2021,
author = {Zhou, Yinong and Liu, Feng},
doi = {10.1021/acs.nanolett.0c03579},
issn = {15306992},
journal = {Nano Lett.},
number = {1},
pages = {230--235},
title = {{Realization of an Antiferromagnetic Superatomic Graphene: Dirac Mott Insulator and Circular Dichroism Hall Effect}},
volume = {21},
year = {2021}
}

@article{Sethi2021,
author = {Sethi, Gurjyot and Zhou, Yinong and Zhu, Linghan and Yang, Li and Liu, Feng},
doi = {10.1103/PhysRevLett.126.196403},
issn = {10797114},
journal = {Phys. Rev. Lett.},
number = {19},
pages = {196403},
publisher = {American Physical Society},
title = {{Flat-Band-Enabled Triplet Excitonic Insulator in a Diatomic Kagome Lattice}},
volume = {126},
year = {2021}
}

@article{Anindya2022,
author = {Anindya, Khalid N. and Rochefort, Alain},
doi = {10.1016/j.cartre.2022.100170},
issn = {26670569},
journal = {Carbon Trends},
month = {apr},
pages = {100170},
publisher = {Elsevier Ltd},
title = {{Controlling the magnetic properties of two-dimensional carbon-based Kagome polymers}},
volume = {7},
year = {2022}
}

@article{Mishra2021c,
author = {Mishra, Shantanu and Catarina, Gon{\c{c}}alo and Wu, Fupeng and Ortiz, Ricardo and Jacob, David and Eimre, Kristjan and Ma, Ji and Pignedoli, Carlo A. and Feng, Xinliang and Ruffieux, Pascal and Fern{\'{a}}ndez-Rossier, Joaqu{\'{i}}n and Fasel, Roman},
doi = {10.1038/s41586-021-03842-3},
issn = {0028-0836},
journal = {Nature},
month = {oct},
number = {7880},
pages = {287--292},
title = {{Observation of fractional edge excitations in nanographene spin chains}},
volume = {598},
year = {2021}
}

@article{Liu2014,
author = {Liu, Zheng and Liu, Feng and Wu, Yong Shi},
doi = {10.1088/1674-1056/23/7/077308},
issn = {16741056},
journal = {Chin. Phys. B},
number = {7},
pages = {077308},
title = {{Exotic electronic states in the world of flat bands: From theory to material}},
volume = {23},
month = {may},
year = {2014}
}

@article{Galeotti2020,
author = {Galeotti, G. and {De Marchi}, F. and Hamzehpoor, E. and MacLean, O. and {Rajeswara Rao}, M. and Chen, Y. and Besteiro, L. V. and Dettmann, D. and Ferrari, L. and Frezza, F. and Sheverdyaeva, P. M. and Liu, R. and Kundu, A. K. and Moras, P. and Ebrahimi, M. and Gallagher, M. C. and Rosei, F. and Perepichka, D. F. and Contini, G.},
doi = {10.1038/s41563-020-0682-z},
journal = {Nat. Mater.},
month = {aug},
number = {8},
pages = {874--880},
publisher = {Nature Research},
title = {{Synthesis of mesoscale ordered two-dimensional $\pi$-conjugated polymers with semiconducting properties}},
volume = {19},
year = {2020}
}

@article{Delgado2023,
title={{Evidence for excitonic insulator ground state in triangulene Kagome lattice}},
journal = {arxiv.org/abs/2301.06171},
author={Aidan Delgado and Carolin Dusold and Jingwei Jiang and Adam Cronin and Steven G. Louie and Felix R. Fischer},
year={2023}
}

@article{Talirz2016,
author = {Talirz, Leopold and Ruffieux, Pascal and Fasel, Roman},
doi = {10.1002/adma.201505738},
journal = {Adv. Mater.},
month = {aug},
number = {29},
pages = {6222--6231},
title = {{On‐Surface Synthesis of Atomically Precise Graphene Nanoribbons}},
volume = {28},
year = {2016}
}

@article{Trunk2017,
author = {Trunk, Matthias and Teichert, Johannes F. and Thomas, Arne},
title = {{Room-Temperature Activation of Hydrogen by Semi-immobilized Frustrated Lewis Pairs in Microporous Polymer Networks}},
journal = {J. Am. Chem. Soc.},
volume = {139},
number = {10},
pages = {3615-3618},
year = {2017},
doi = {10.1021/jacs.6b13147}
}

@article{Horcas2007,
author = {Horcas, I. and Fernández, R. and Gómez-Rodríguez, J. M. and Colchero, J. and Gómez-Herrero, J. and Baro, A. M.},
title = {{WSXM: A software for scanning probe microscopy and a tool for nanotechnology}},
journal = {Rev. Sci. Instrum.},
volume = {78},
number = {1},
pages = {013705},
year = {2007},
month = {01},
doi = {10.1063/1.2432410}
}

@article{Yu2024,
author = {Yu, Hongde and Heine, Thomas},
doi = {10.1126/sciadv.adq7954},
journal = {Sci. Adv.},
month = {oct},
number = {40},
pages = {eadq7954},
title = {{Prediction of metal-free Stoner and Mott-Hubbard magnetism in triangulene-based two-dimensional polymers}},
volume = {10},
year = {2024}
}

@article{Li2021,
author = {Li, Jingcheng and Sanz, Sofia and Merino-D{\'{i}}ez, Nestor and Vilas-Varela, Manuel and Garcia-Lekue, Aran and Corso, Martina and de Oteyza, Dimas G. and Frederiksen, Thomas and Pe{\~{n}}a, Diego and Pascual, Jose Ignacio},
doi = {10.1038/s41467-021-25688-z},
journal = {Nat. Commun.},
month = {sep},
number = {1},
pages = {5538},
title = {{Topological phase transition in chiral graphene nanoribbons: from edge bands to end states}},
volume = {12},
year = {2021}
}

@article{Kresse1996,
author = {Kresse, G. and Furthm\"uller, J.},
journal = {Phys. Rev. B},
volume = {54},
issue = {16},
pages = {11169--11186},
title = {{Efficient iterative schemes for ab initio total-energy calculations using a plane-wave basis set}},
numpages = {0},
month = {Oct},
publisher = {American Physical Society},
doi = {10.1103/PhysRevB.54.11169},
year = {1996}
}

@article{Perdew1996,
author = {Perdew, John P. and Ernzerhof, Matthias and Burke, Kieron},
title = {{Rationale for mixing exact exchange with density functional approximations}},
journal = {J. Chem. Phys.},
volume = {105},
number = {22},
pages = {9982-9985},
month = {12},
issn = {0021-9606},
doi = {10.1063/1.472933},
year = {1996}
}

@article{Ernzerhof1999,
author = {Ernzerhof, Matthias and Scuseria, Gustavo E.},
title = {{Assessment of the Perdew–Burke–Ernzerhof exchange-correlation functional}},
journal = {J. Chem. Phys.},
volume = {110},
number = {11},
pages = {5029-5036},
month = {03},
issn = {0021-9606},
doi = {10.1063/1.478401},
year = {1999}
}

@article{Adamo1999,
author = {Adamo, Carlo and Barone, Vincenzo},
title = {{Toward reliable density functional methods without adjustable parameters: The PBE0 model}},
journal = {J. Chem. Phys.},
volume = {110},
number = {13},
pages = {6158-6170},
month = {04},
issn = {0021-9606},
doi = {10.1063/1.478522},
year = {1999}
}

\end{document}